\documentclass[runningheads]{llncs}
\usepackage[T1]{fontenc}
\usepackage{cite}
\usepackage{amsmath,amssymb,amsfonts}
\usepackage{algorithmic}
\usepackage{graphicx}
\usepackage{textcomp}
\usepackage{xcolor}
\usepackage{url}
\usepackage{caption}

\usepackage{array}
\usepackage{multirow}
\usepackage{soul}

\usepackage[utf8]{inputenc}

\usepackage{graphicx}
\usepackage{subcaption} 

\usepackage{booktabs} 
\usepackage{geometry} 
\newcolumntype{L}[1]{>{\raggedright\arraybackslash}p{#1}} 
\newcolumntype{R}[1]{>{\raggedleft\arraybackslash}p{#1}}  
\newcolumntype{C}[1]{>{\centering\arraybackslash}p{#1}}   

\begin{document}
\title{Revolutionizing Bacterial Genomics: Graph-Based Strategies for Improved Variant Identification}

\author{Fathima Nuzla Ismail\inst{1} \orcidID{0000-0002-0716-1478} \and Abira Sengupta\inst{2} \orcidID{0000-0002-6867-3362}}
\institute{Dept. of Mathematics,State University of New York at Buffalo, USA\\
\email{fathima.nuzla.ismail@gmail.com} \and School of Computing, University of Otago, New Zealand,\\
\email{enggabira0609@gmail.com}}

\maketitle              
%



\begin{abstract}
A significant advancement in bioinformatics is using genome graph techniques to improve variation discovery across organisms. Traditional approaches, such as bwa mem, rely on linear reference genomes for genomic analyses but may introduce biases when applied to highly diverse bacterial genomes of the same species. Pangenome graphs provide an alternative paradigm for evaluating structural and minor variations within a graphical framework, including insertions, deletions, and single nucleotide polymorphisms. Pangenome graphs enhance the detection and interpretation of complex genetic variants by representing the full genetic diversity of a species. In this study, we present a robust and reliable bioinformatics pipeline utilising the PanGenome Graph Builder (PGGB) and the Variation Graph toolbox (vg giraffe) to align whole-genome sequencing data, call variants against a graph reference, and construct pangenomes from assembled genomes. Our results demonstrate that leveraging pangenome graphs over a single linear reference genome significantly improves mapping rates and variant calling accuracy for simulated and actual bacterial pathogens datasets.

\keywords{Pangenome graphs \and Structural variations \and Graph reference genomes \and Variant calling}
\end{abstract}

%
%
%

\section{Introduction}

Neisseria meningitidis is a gram-negative, encapsulated bacterium responsible for meningococcal disease, including meningitis and septicemia. It is a leading cause of bacterial meningitis, particularly affecting infants, adolescents, and young adults \cite{stephens2007epidemic}. The bacterium’s genome exhibits significant diversity due to horizontal gene transfer and recombination, through which genetic material is exchanged between different organisms. This genetic variability complicates vaccine development and necessitates continuous genomic surveillance \cite{Maiden:1998}.

\begin{figure}[!htbp]
    \centering
    \vspace{-10pt} 
    \includegraphics[width=\columnwidth]{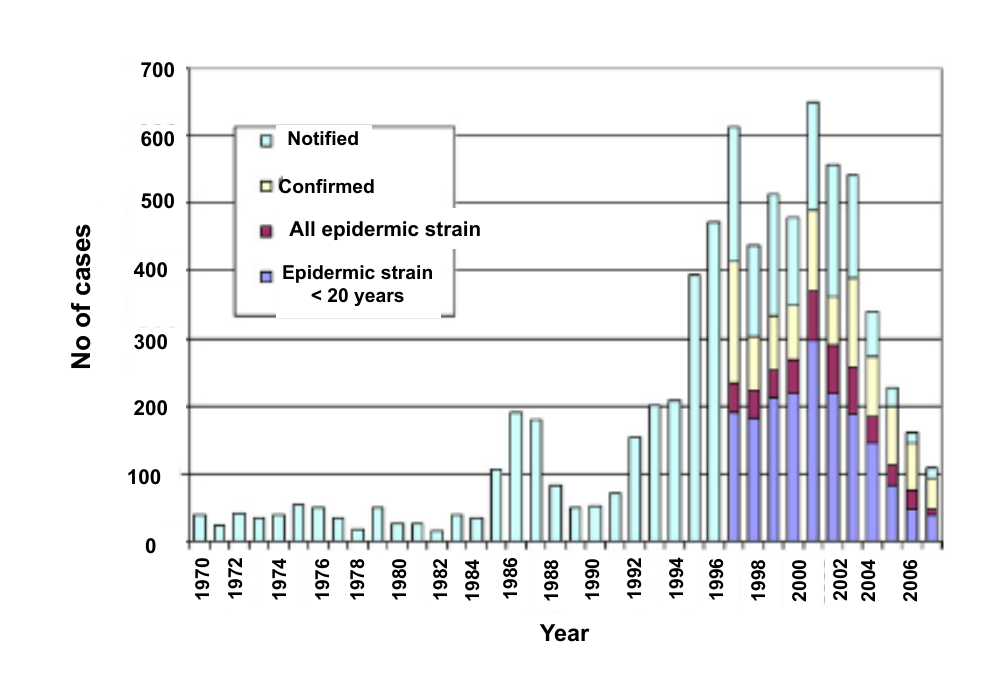} 
    \caption{Meningococcal disease in New Zealand 1970–2007 \cite{loring2008menzb}.}
    \label{fig:fig-graph}
\end{figure}

Ongoing research on N. meningitidis is crucial for public health, particularly in New Zealand, where a serogroup B epidemic occurred from 1991 to 2007, and the distribution of this epidemic spread across New Zealand is visualized in Figure~\ref{fig:fig-graph}. This research has revealed extensive genetic variability among isolates, informing vaccination strategies like the MeNZB vaccine. However, cases involving other serogroups still arise, highlighting the need for ongoing surveillance. Globally, N. meningitidis continues to cause significant morbidity and mortality, especially in sub-Saharan Africa’s ``meningitis belt'', which stretches from Senegal in the west to Ethiopia in the east. The disease is endemic in this region due to specific environmental and socio-economic factors \cite{greenwood1999manson}. Emerging hypervirulent clones have been identified through whole-genome sequencing (WGS), which facilitates real-time antimicrobial resistance monitoring and aids in vaccine development \cite{parkhill2000complete}. Genomic research has greatly advanced our understanding of N. meningitidis, which is essential for managing meningococcal disease in New Zealand. This fact underscores the vital role of public health professionals, researchers, and policymakers in strengthening public health systems worldwide. 

Conventional genomic analyses that rely on a single linear reference genome have certain drawbacks, mainly when dealing with highly variable genomes like those of RNA viruses and bacteria. Such methods often overlook accessory genome changes and structural variants (SVs), resulting in an inadequate understanding of pathogen diversity and evolution \cite{Darmon:2014, Eizenga:2020}. Viral genomes present additional challenges due to their high mutation rates and capacity for interspecies transmission \cite{Sanjuan:2016, Domingo:2021}.

Pangenome graphs offer a promising solution in genomic research. By compactly depicting entire genomes, their homologies, and all variations between them, pangenome graphs allow researchers to study sequence evolution and variation across various species. These graphs help identify recombination events, measure conservation, detect variation, and infer relationships between species. They have become a revolutionary tool for examining genomic variation and addressing these challenges. By integrating different forms of variation, such as single nucleotide polymorphisms (SNPs), SVs, and rearrangements, pangenome graphs provide a comprehensive framework for genomic comparison, avoiding the biases inherent in single-reference methods. The potential of pangenome graphs is immense and is fostering optimism for the future of genomic research.

By evenly aligning several genomes, programs such as PanGenome Graph Builder (PGGB) (Section~\ref{PGGB}) offer an objective way to create these graphs, guaranteeing an accurate depiction of every variant and instilling confidence in the research methods \cite{Garrison:2023}. 


The highly recombinogenic genome of Neisseria meningitidis, the bacterium responsible for septicaemia and meningitis, was analysed in this study using PGGB with three types of datasets: the Neisseria meningitidis dataset, the original reference genome (NC\_017518), and a simulated dataset.

A recombinogenic genome has a high rate of recombination, a process where genetic material is exchanged between different organisms or different regions of the same organism's DNA. Better studies of outbreak dynamics, treatment resistance, and vaccine design were made possible by the pipeline's increased variant calling and mapping rates \cite{Rasko:2008, Naz:2019, Yang:2021}. This highlights how the study of pathogenic diseases may transform by using pangenome graphs, which offer thorough insights into pathogen diversity, evolution, and genomic architecture.  

Despite the challenges posed by highly varied genomes, the advancements in WGS have significantly transformed infectious disease surveillance \cite{Didelot:2012, Holt:2015}. Pangenome graphs consistently depict genomic variation and reduce biases in conventional methods \cite{Paten:2017, Garrison:2018}. Tools such as PGGB demonstrate the effectiveness of graph-based techniques in thorough genetic investigations, especially for recombinogenic diseases like \emph{N. meningitidis} \cite{Garrison:2023}. This paradigm shift, which refers to a fundamental change in the way we approach pathogen genomics and public health interventions, underscores the importance of graph-based approaches. As this study shows, these methods offer a thorough and reliable understanding of pathogen variety and evolution, which improves our ability to combat infectious diseases. 

The paper's first section describes the genomic complexity of the highly recombinogenic bacterium Neisseria meningitidis, which causes meningococcal illness. It draws attention to the inability of linear reference-based genome analysis to fully capture genetic variability. Conventional tools like bwa mem fail to represent structural variants (SVs) and accessory genomic elements in highly diverse bacterial genomes. We aim to implement and evaluate a graph-based genomic pipeline that improves variant detection and reduces reference bias.


\section{Materials and methods}


\subsection{Use cases}\label{ProA}

\textbf{Usecase-1}: Housekeeping genes in N. meningitidis isolates were classified into three phylogenetic clades clade 154, clade 41, and clade 42—based on unpublished whole-genome sequencing (WGS) data \cite{Maiden:1998}. The epidemic was primarily driven by two monophyletic clades: ST154 and ST42. The reference genome NC\_017518, derived from an ST42 isolate, was utilised to analyse the WGS dataset. Nanopore long-read sequencing was employed to generate complete reference genomes for NMI01191 (an ST41 isolate) and NMI97348 (an ST154 isolate) using sequencing libraries prepared with 400 ng of high molecular weight genomic DNA. 

De novo assembly was performed using Flye version 2.8.1 \cite{Kolmogorov:2019}, while Unicycler version 0.4.8 \cite{Wick:2017} was used for assembly polishing with Illumina sequencing data. SimuG, a lightweight simulation tool, was also employed to generate genomic sequences with predefined random variants based on reference sequences \cite{Yue:2019}. Pangenome graphs are constructed using simulated genomic data with different mutations (SNPs, INDELs, CNVs, and INVs) and three clade-specific genome assemblies (ST154, ST41, and ST42). Three distinct simulated sequences were created: the first contained 4000 single nucleotide polymorphisms (SNPs) and 400 insertions/deletions (INDELs), maintaining an insertion-to-deletion ratio of 1:4; the second had the same SNP and INDEL counts plus four genomic inversions; and the third included 4000 SNPs, 400 INDELs, and four copy number variations (CNVs). These simulated sequences were used to construct a genomic variation graph, enabling an evaluation of the accuracy of graph-based analyses in detecting SNPs, INDELs, inversions, and CNVs. This approach provides valuable insights into the robustness and efficacy of graph-based methods for genomic variant detection.


\begin{figure*}[!htbp]
    \centering
   \vspace{-10pt} 
    \includegraphics[width=1\textwidth]{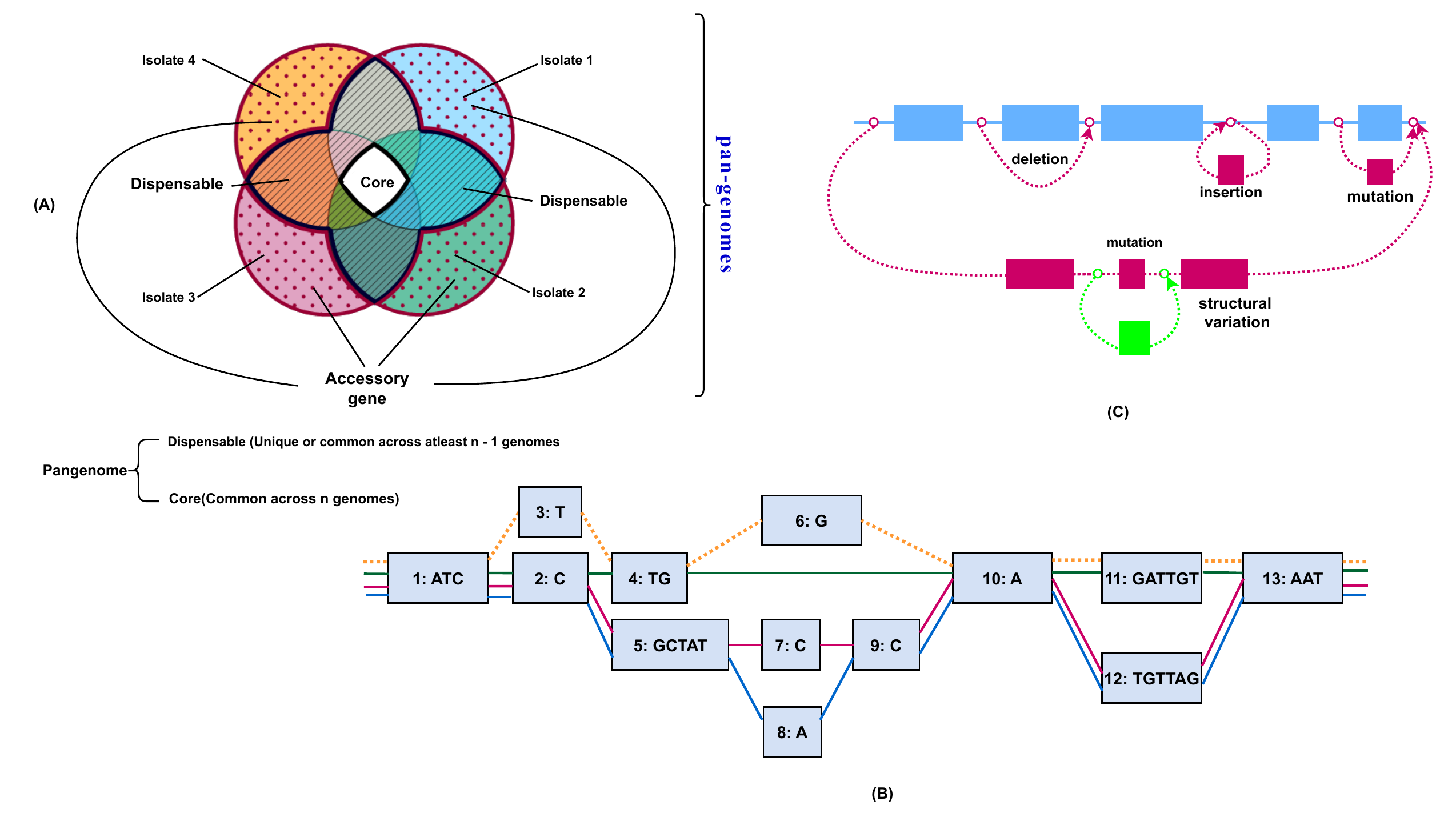} 
    \caption{The era of pangenomes; (A) represents the pangenome, core genome, and accessory genomes. The pangenome represents the comprehensive collection of whole-genome sequences from several individuals within a clade. Core genomes are constructed up of genes that are present in all individuals, whereas accessory genomes consist of genes found in a subset of individual genomes. (B) a representation of pangenome graphs. Pangenomes are represented by graph models, in which DNA segments of different lengths are represented by nodes, which are numerically labelled. Edges connect nodes, and paths represent walks through the nodes of the graph, which correspond to the input genomes. (C) graph-based representation of pangenome.}
    \label{fig:fig1}
\end{figure*}

\textbf{Usecase-2}: Accurate variant calling serves as the foundation of genomic research, pivotal in uncovering genetic variations with critical implications for understanding diseases, advancing diagnostics, and developing therapeutic strategies. While linear reference-based alignment methods like bwa mem have long been the standard for variant detection, they often struggle to capture the complete range of genetic diversity, particularly in complex genomic regions containing structural variations. In contrast, graph-based approaches such as vg giraffe offer a promising solution by utilising pangenome graphs to provide a more comprehensive representation of genetic variation.

In this study, as the second analysis, we compare the graph-based approach (vg giraffe) performance with the linear method (bwa mem) for variant calling in a simulated dataset. Variant calling with vg giraffe versus bwa mem, validating results by comparing them to ground-truth datasets. The tools used for Graph Construction were PGGB for graph construction with gfaffix, smoothxg, seqwish, and wfmash. GFAESTUS and ODGI were used for analysis and visualization—VG toolset for evaluating and extracting variants. Our results demonstrate a clear advantage of the graph-based framework in terms of sensitivity, specificity, and accuracy.


\begin{figure}
    \centering
    \includegraphics[width=0.9\linewidth]{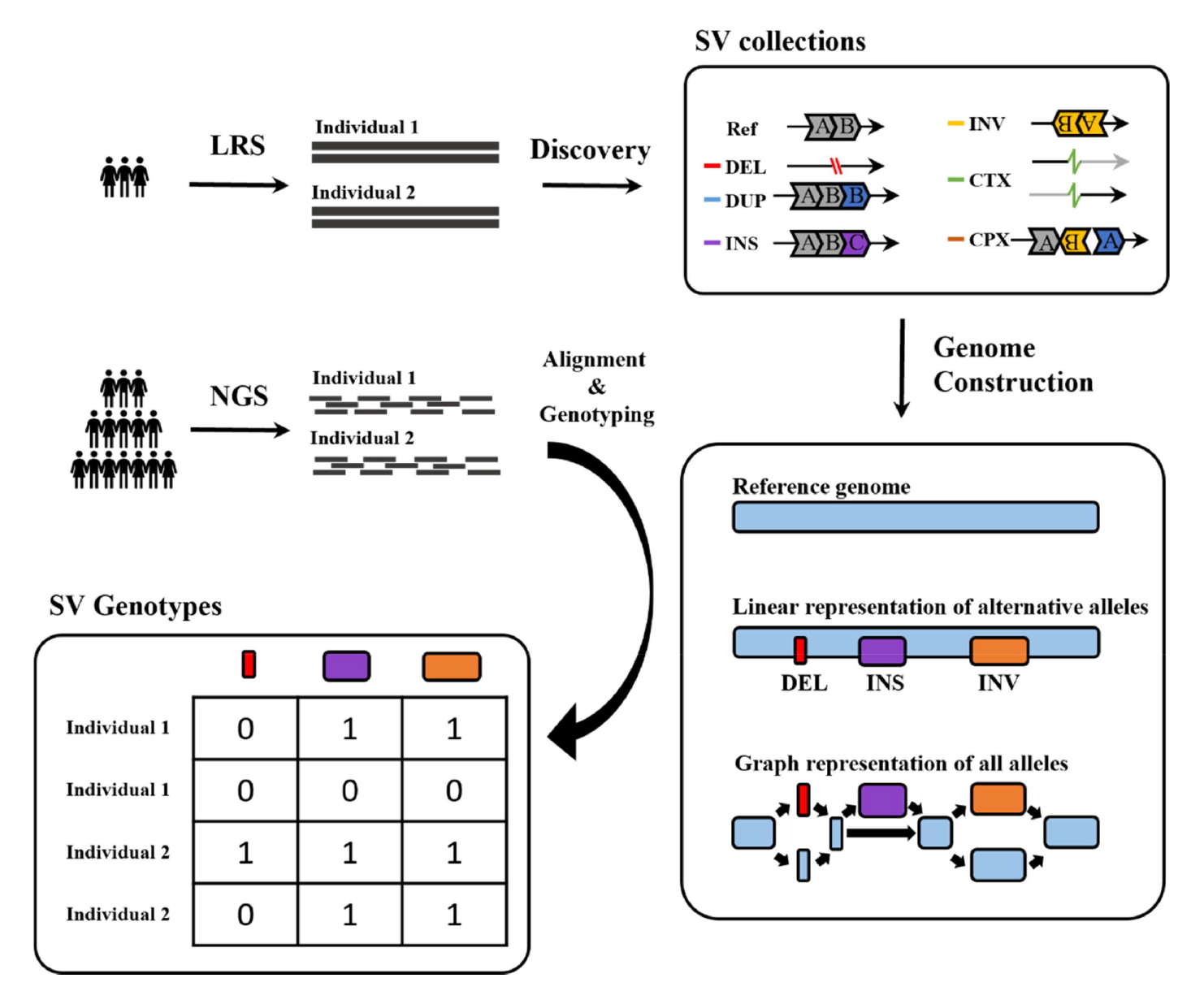}
    \caption{Fig. 5 An illustration of the hybrid sequencing approach, a practical method that can be applied in various research settings. Initially, a small set of long-read sequencing (LRS) samples is used for variant discovery, followed by a more extensive set of short-read sequencing (SRS) samples for genotyping. Structural variant (SV) collections, including deletions (DEL), duplications (DUP), insertions (INS), inversions (INV), inter-chromosomal translocations (CTX), and complex SVs (CPX), are identified and integrated into the reference. These variants are then used to construct either a linear representation of alternative alleles or a graph representation encompassing all alleles. Genotyping uses two strategies: aligning short reads to the primary contig with alternative sequences or conducting sequence-to-graph alignment.}
    \label{fig:enter-label-1}
\end{figure}

\section{Tools: The PanGenome Graph Builder-PGGB, ODGI, VG Toolkit}\label{PGGB}

\textbf{We introduce a Graph-Based Variant Calling Pipeline:} the development of a novel pipeline using PGGB (PanGenome Graph Builder) and the VG toolkit (vg giraffe) to construct and utilise pangenome graphs. Also, we benchmark against traditional linear methods using actual and simulated datasets, enhancing F1 scores, sensitivity, and specificity when employing the graph-based method, particularly in intricate genomic areas.

\begin{figure*}[!htbp]
    \centering
   \vspace{-10pt} 
    \includegraphics[width=0.95\textwidth]{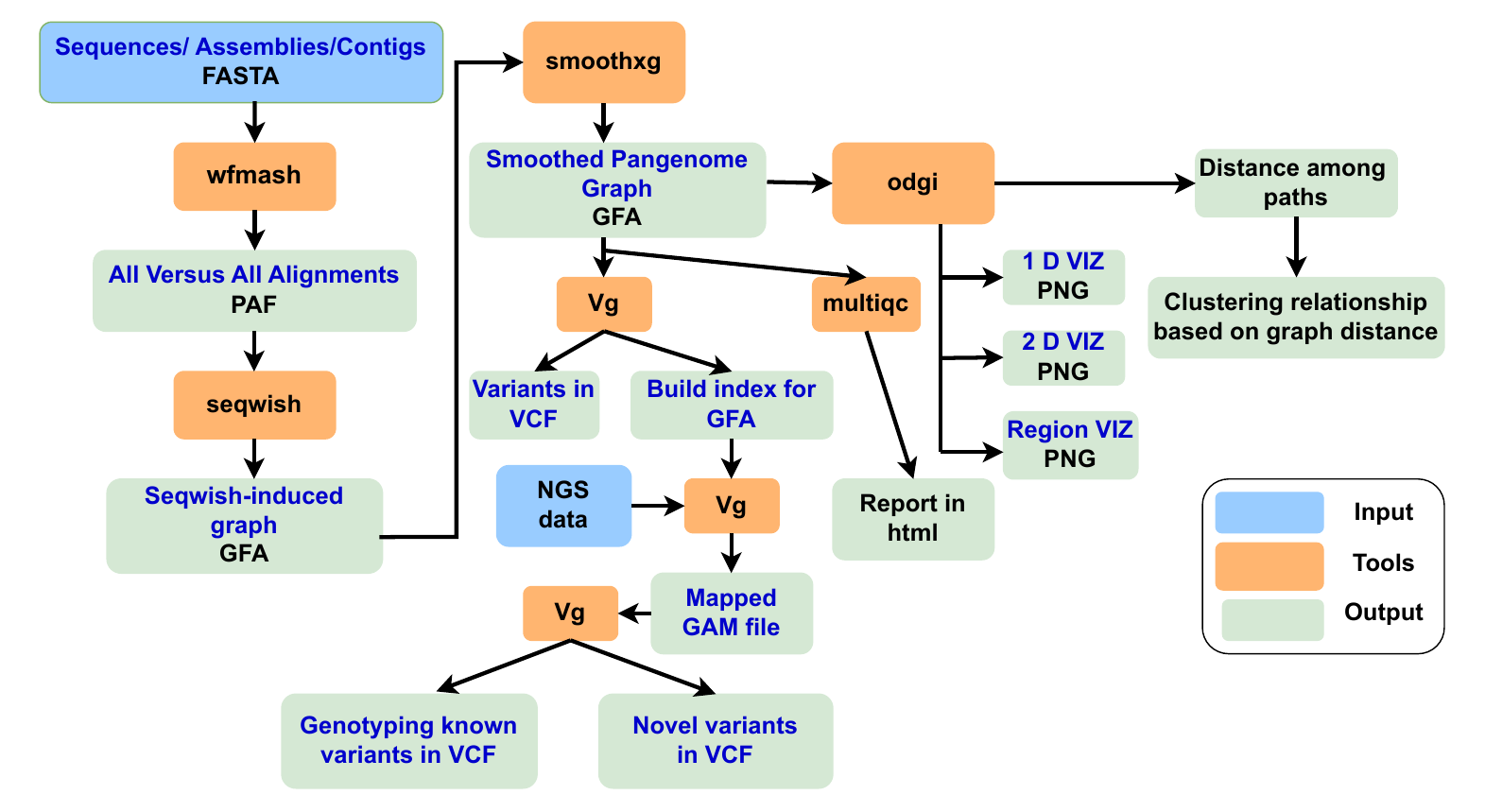} 
    \caption{PGGB pangenome graph pipeline. The pipeline consists of variant calling for NGS data using the VG toolkit, graph construction using the PGGB tool, and graph manipulation using the Optimised Dynamic Genome/Graph Implementation (ODGI) tool. The overview shows how pangenome analysis can be done effectively and comprehensively.}
    \label{fig:fig2}
\end{figure*}

The PanGenome Graph Builder (PGGB) was used to construct pangenome graphs for the ST42Simulated- Data, ST41SimulatedData, and ST154SimulatedData genomes (Figure~\ref{fig:fig2}). PGGB employs an all-to-all alignment approach, utilising tools such as wfmash and seqwish for graph generation and smoothxg and gfaffix for progressive graph normalisation. Statistical analysis and visualisation were performed using the Optimised Dynamic Genome/Graph Implementation (ODGI) toolkit \cite{Guarracino:2022, Garrison:2023}. Before building the graphs, genome alignment was optimised to ensure uniform starting points in the sequences, with Circlator v1.5.5 used for alignment to the dnaA gene \cite{Hunt:2015}.

 Parameter tuning was guided by pairwise distances calculated using the mash triangle approach \cite{Ondov:2016}. For comprehensive alignments, the parameter \texttt{-p} was set below the largest pairwise distances, and the parameter \texttt{-k} was adjusted to exclude matches shorter than a specified length. This adjustment aimed to minimise complexity due to short homologies, such as those introduced by transposable elements.

ODGI statistics were employed to evaluate the resulting pangenome graphs, which were visualised in two dimensions using gfaestus. Graph statistics were compiled and presented using MultiQC. All computational analyses were performed on a high-performance server equipped with 49 threads, a 6200 CPU, and 4TB of RAM, enabling efficient data processing. The systematic adjustment of parameters resulted in well-constructed and informative pangenome graphs, effectively balancing alignment precision with graph complexity. These findings highlight the utility of PGGB for diverse datasets in infectious disease and comparative genomics research \cite{Garrison:2023}.

Variation graphs, which provide a compact and bidirectional representation of genetic variation across populations, were utilised to capture both small and large structural variants (SVs), such as inversions and duplications. The Variation Graph (VG) toolkit played a crucial role by extracting genetic variants from pangenome graphs into variant call format (VCF) files along the ST42 path. VG is known for its robust genomic analysis capabilities, including alignment, assembly, and variant calling, while maintaining the positional context of genomic variations through graph-based structures.

The VG deconstruct function was utilised to extract variants from the graph, ensuring that all snarls, graph paths, and conflicting genotypes were included. Variants were compared against simulated genomes (ST42SimulatedData, ST41SimulatedData, and ST154SimulatedData) derived from ST42, which served as the ground truth. Variants larger than 100 bp were analysed directly, while smaller variants were processed using vcfallelicprimitives from vcflib to resolve complex structures. This method enabled the accurate classification of genetic variants, supporting the study's objective of evaluating graph-based genomic approaches.


Simulated genomes based on ST42 were ground truth to validate the variant identification process. Variants more significant than $100$ bp were directly analysed, while smaller ones were simplified using vcfallelicprimitives. Detected variants were compared against ground truth, revealing consistent, false negative, and false positive classifications. This method assessed the accuracy and reliability of variation graphs in capturing genomic diversity.

Using the graph's routes for comparisons, variant calling with the VG toolkit differed from linear reference-based approaches. The genotype (GT) and PASS columns indicated variants according to their confidence levels. Variants with PASS but GT $\neq$ 1|1 were categorised as errors, whereas those with PASS and GT = 1|1 were categorised as high-confidence. Although the ST41 and ST154 groups had lower percentages because of their higher genomic variety and lack of specific references in the graph, simulated datasets consistently displayed high-confidence variations. Adding these group-specific references enhanced the graph's high-confidence variant proportions.

\begin{figure*}[!htbp]
    \centering
   \vspace{-10pt} 
    \includegraphics[width=1\textwidth]{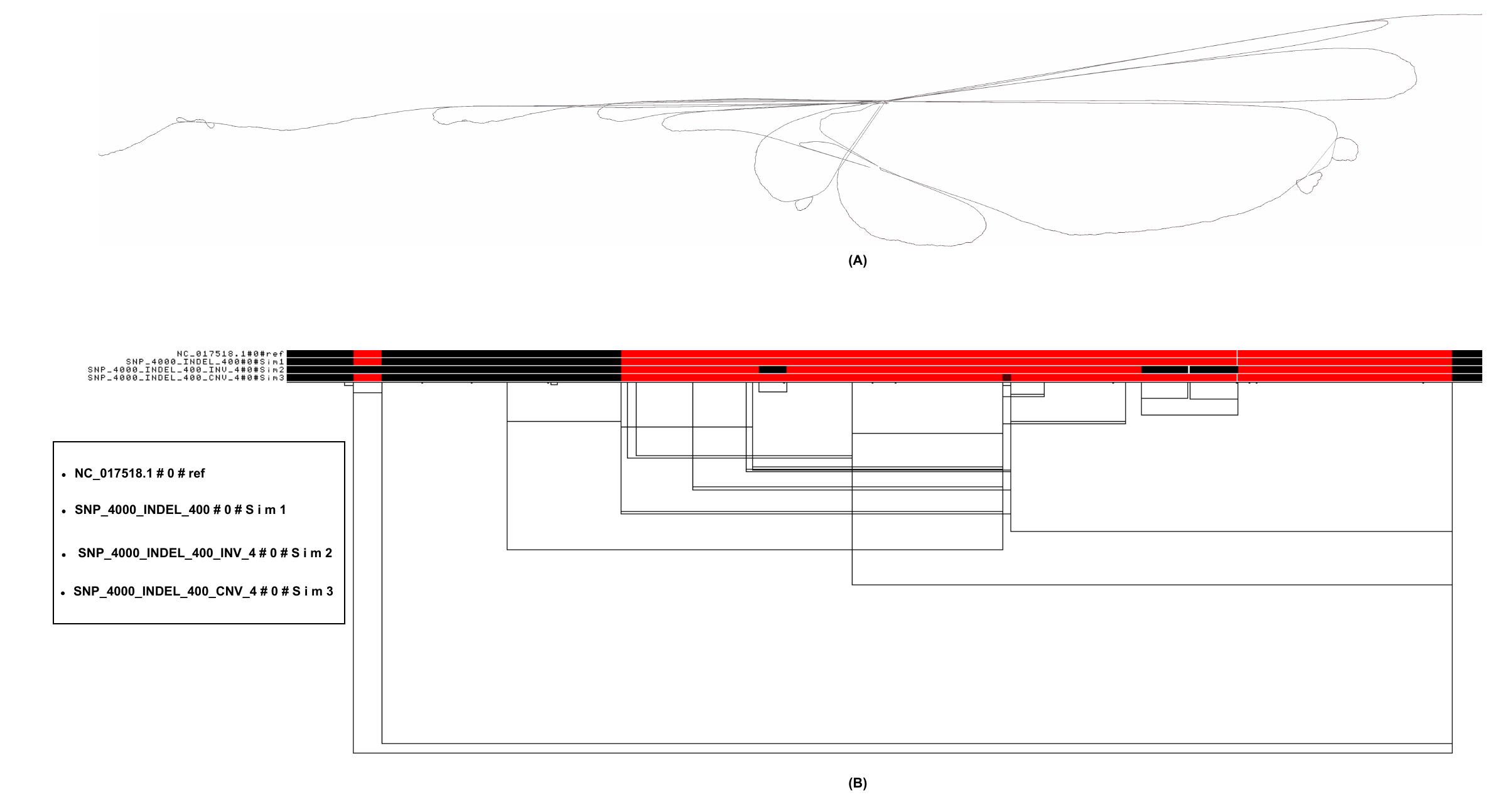} 
    \caption{(A) Two-dimensional visualisation with the ODGI 1D (B) ODGI 1D visualisation by path orientation. Black and red indicate that the path traverses the graph in the forward and reverse orientation, respectively. }
    \label{fig:fig3}
\end{figure*}

The visual outputs from the pangenome graph tool are comprehensive and go with the in-depth variant statistics offered by the VG tools stats analysis. 
The alignment of genomic sequences and their variants, shown in the Figure~\ref{fig:fig3} (A), clearly illustrates the pangenome graph structure. The linear segments in the graph correspond to conserved regions shared across all genomes, while the branching paths represent areas of genetic variation. These branches align with the identified SNPs ($3,983$), INDELs ($400$), and more significant structural variants such as inversions (INVs) and copy number variations (CNVs). Specifically, the dataset includes 4 INVs, each at least $50,000$ bases, and $4$ CNVs, each $20,000$ bases, represented by significant branching patterns in the graph. This visualisation underscores the graph's ability to integrate fine-scale variations and more significant, complex structural rearrangements.

The Figure~\ref{fig:fig3} (B) represents the block alignment structure of the pangenome. Each row corresponds to a genome sequence, such as the reference and simulated data sets (Sim1, Sim2, Sim3), and highlights regions of similarity and divergence. The black blocks signify highly conserved regions with minimal variation, aligning with the shared genomic backbone observed in the graph. In contrast, the red blocks represent divergent regions consistent with the detected SNPs, INDELs, INVs, and CNVs. The overlapping lines and boxed areas emphasise structural rearrangements like inversions and copy number changes, further highlighting the structural diversity captured in the analysis.


\section{Results - linear (bwa mem) vs graphical (vg giraffe) tool}\label{Results}

The results from the variant calling comparison between the linear method (bwa mem) and the graph method (vg giraffe) indicate a clear advantage for the graph-based approach. In the report generated
for vg giraffe, the method identified 5,726 True Positives (TP), compared to 5,328 TP for bwa mem. This difference in True Positives underscores the superior effectiveness of the VG giraffe in detecting variants in the simulated dataset. Furthermore, the vg giraffe demonstrated a Sensitivity of 95.43\%, significantly outperforming bwa mem’s 88.80\%.
This means that VG Giraffe was better at correctly identifying true variants in the ground truth data. Moreover, the Specificity of the vg giraffe was also superior at 99.88\%, compared to bwa mem’s 99.87\%.
This indicates that the VG giraffe detects more variants and maintains a high accuracy rate in distinguishing true negatives from false positives. The F1 Score, which balances precision and recall, was higher for vg giraffe at 79.98\%, compared to 74.46\% for bwa mem (Table~\ref{table:vg}). These results suggest that the graph-based approach using vg giraffe provides a more robust and accurate framework for variant calling than the traditional linear method employed by bwa mem. This improved performance is especially advantageous in complex genomic regions with structural variations, establishing VG giraffe as a superior option for variant detection in genomic research.

Figure~\ref{fig:fig3} (A) and (B) and the VG tools statistics show the accuracy and strength of pangenome graphs in identifying genomic variants. Pangenome graphs avoid the drawbacks of conventional linear references by utilising graph-based genomic representations, especially when detecting complicated genomic rearrangements and structural variations.
The following was the parameter adjustment for each dataset: 

\begin{enumerate}

    \item \textbf{Consistent:} consistent presence in both the graph and ground truth.
    
    \item \textbf{False negatives:} false negatives found in the ground truth but missed in the graph.
    
    \item \textbf{False positives:} A false positive is detected in the graph but absent in the ground truth.
    
\end{enumerate}
\begin{itemize}

\item \textbf{True Positive (TP)} = SNP + INDELs which are exactly matched in groud\_truth.vcf and simulated.vcf (TP = 4, 993 + 733 = 5,726)

\item \textbf{False Positive (FP)} = SNPs + INDELs private to simulated.vcf and not found in groud\_truth.vcf (FP = 1,839 + 753 = 2,592).

\item \textbf{True Negative (TN)} = Length of Reference the Sequence - Ground Truth SNPs - Ground Truth INDELs - False Positive. (TN = 2,248,966 - 5,000 - 1,000 - 2,592 = 2,240,374).

\item \textbf{False Negative (FN)} = Ground Truth SNPs + Ground Truth INDELs - True Positive. (FN=5,000+1,000-5,726=274). 

\item \textbf{Sensitivity}, \textbf{Specificity}, and \textbf{F1 Score} will be,
\end{itemize}


\begin{equation}
\text{Sensitivity} = \frac{TP}{TP + FN} 
= \frac{5,726}{5,726 + 274} 
= 95.4333\%
\end{equation}


\begin{equation}
\text{Specificity} = \frac{TN}{TN + FP} \\
= \frac{2,240,374}{2,240,374 + 2,592} \\
= 99.8844\%
\end{equation}


\begin{equation}
\begin{aligned}
\text{F1 score} &= \frac{TP}{TP + \frac{1}{2} (FP + FN)} \\
&= \frac{5,726 }{5,726  + 0.5 \times (2,592 + 274)} \\
&= 79.9832\%
\end{aligned}
\end{equation}




\begin{table}[htbp]
\caption{Comparison between linear (bwa mem) vs graphical (vg giraffe) tool}
\begin{center}
\resizebox{\columnwidth}{!}{%
\renewcommand{\arraystretch}{1.5}
\begin{tabular}{|l|r|r|r|r|r|r|r|}
\hline
\textbf{Method} &  \textbf{TP} &   \textbf{TN}  &   \textbf{FP}  &  \textbf{FN} & \textbf{Sensitivity} &  \textbf{Specificity} &  \textbf{F1 Score}\\
\hline
bwa mem (Linear)   & 5,328 &  2,239,982 & 2,984 & 672 & 88.8000\%  &   99.8669\%  & 74.4550\% \\
\hline
vg giraffe (Graph) & 5,726 &  2,240,374 &  2,592  &  274  &    95.4333\% & 99.8844\%  &  79.9832\%  \\
\hline
\end{tabular}%
}
\label{table:vg}
\end{center}
\end{table}



\section{Results-Variant Mapping Rate using Simulated Data PGGB}

ST42 is the graph's first path, which impacts variant calling accuracy. Reflecting the influence of path selection on variant calling outcomes, ST42 and ST42SimulatedData had the highest ratios of high-confidence variants to simulated variants ($0.944–0.9706$), whereas ST154Mutation and ST154SimulatedData had comparatively lower ratios ($0.8755–0.9049$). This investigation underscores the profound significance of leveraging pangenome graph references for precise mapping, genotyping, and variant calling in a diverse array of complex genomic datasets.

Mapping rates significantly increase when next-generation sequencing (NGS) data is analyzed using a pangenome graph reference. In comparison to linear references (e.g., ST42, ST42SimulatedData, ST41SimulatedData, and ST154SimulatedData), which had mapping rates above $99$\% but showed bias, simulated NGS datasets based on the \emph{4SimulatedData} pangenome graph achieved a $100$\% mapping rate when mapped to the graph (Figure~\ref{fig:fig2}). This demonstrates the superior handling of genetic diversity (the variety of genetic information within a species or population) by pangenome graph references, with improved inclusivity and accuracy, thereby enhancing the quality of NGS data analysis. 

The Graph-based method not only enables efficient variant genotyping but also ensures practicality, providing a user-friendly approach to genomic research. The ST42SimulatedData group had the most variants in NGS datasets, consistent with its simulation of $4000$ SNPs and $400$ indels. The results were consistent across three simulated datasets. The graph's ability to better capture genetic variations is demonstrated by the fact that the ST41SimulatedData and ST154SimulatedData groups found more variants than their respective linear references (Figure~\ref{fig:fig2})). 


\begin{table}[htbp]
\caption{According to the index, there are three samples in this assembly, which are detailed in the table.}
\begin{center}
\renewcommand{\arraystretch}{1.5}
\begin{tabular}{|L{0.25\columnwidth}|R{0.15\columnwidth}|R{0.09\columnwidth}|R{0.08\columnwidth}|R{0.07\columnwidth}|R{0.07\columnwidth}|}
\hline
\textbf{Name} & \textbf{Length} &\textbf{SNPs} & \textbf{INDELs} & \textbf{INV} & \textbf{CNV} \\
\hline
NC\_017518.1 (Reference) & $2,248,966$ & N/A &	N/A & N/A &	N/A\\
\hline
NC\_017518.1 \_SNP\_4000 \_INDEL\_400 & $2,248,375$ & $4,000$ & $4,00$ & $0$ & $0$\\
\hline
NC\_017518.1 \_SNP\_4000\_INDEL \_400\_INV\_4 & $2,248,375$ & $4,000$	& $4,00$ & $4$ & $0$\\
\hline
NC\_017518.1 \_SNP\_4000\_INDEL \_400\_CNV\_4 & $2,499,008$ &	$4,000$ & $4,00$ & $0$ & $4$\\
\hline
\end{tabular}
\label{tab:Tab1}
\end{center}
\end{table}


The smaller range of mapping rate variation suggests that the isolates of the ST154 group may be less diverse. In contrast, the more extensive ranges of mapping rate variation ($0.958$ to $0.9956$, $0.971$ to $1$, $0.9744$ to $0.9988$, and $0.9785$ to $0.9998$, respectively) suggest that the isolates of the ST41 group exhibit greater diversity. When mapped to the ST42 and 3ST genome graphs, the isolates in the ST42 group showed similar mapping rates. When these isolates were mapped to ST154 ($0.9536$ to $0.994$) and ST41 ($0.9559$ to $0.9959$), somewhat lower mapping rates were noted. In conclusion, these results underscore the importance of considering multiple references, as they raise the possibility of reference bias when relying just on one linear reference. Nevertheless, the findings suggest that using a graph as a reference can improve the mapping process and minimise the potential for reference bias associated with linear references, providing valuable insights for future research in this field.

\paragraph{\textbf{Structural Variant Analysis and Visualization:}} SNPs, INDELs, INVs, and CNVs were all recorded on the graph via ODGI visualizations. Proven capacity to depict both large-scale and fine-grained structural changes in this research. Statistical comparisons were made for parameter modifications, and true/false positive/negative rates quantitatively supported the superiority of graph-based approaches.





\begin{figure}[ht]
    \centering
    \begin{subfigure}[b]{0.42\linewidth}
        \centering
        \includegraphics[width=\linewidth]{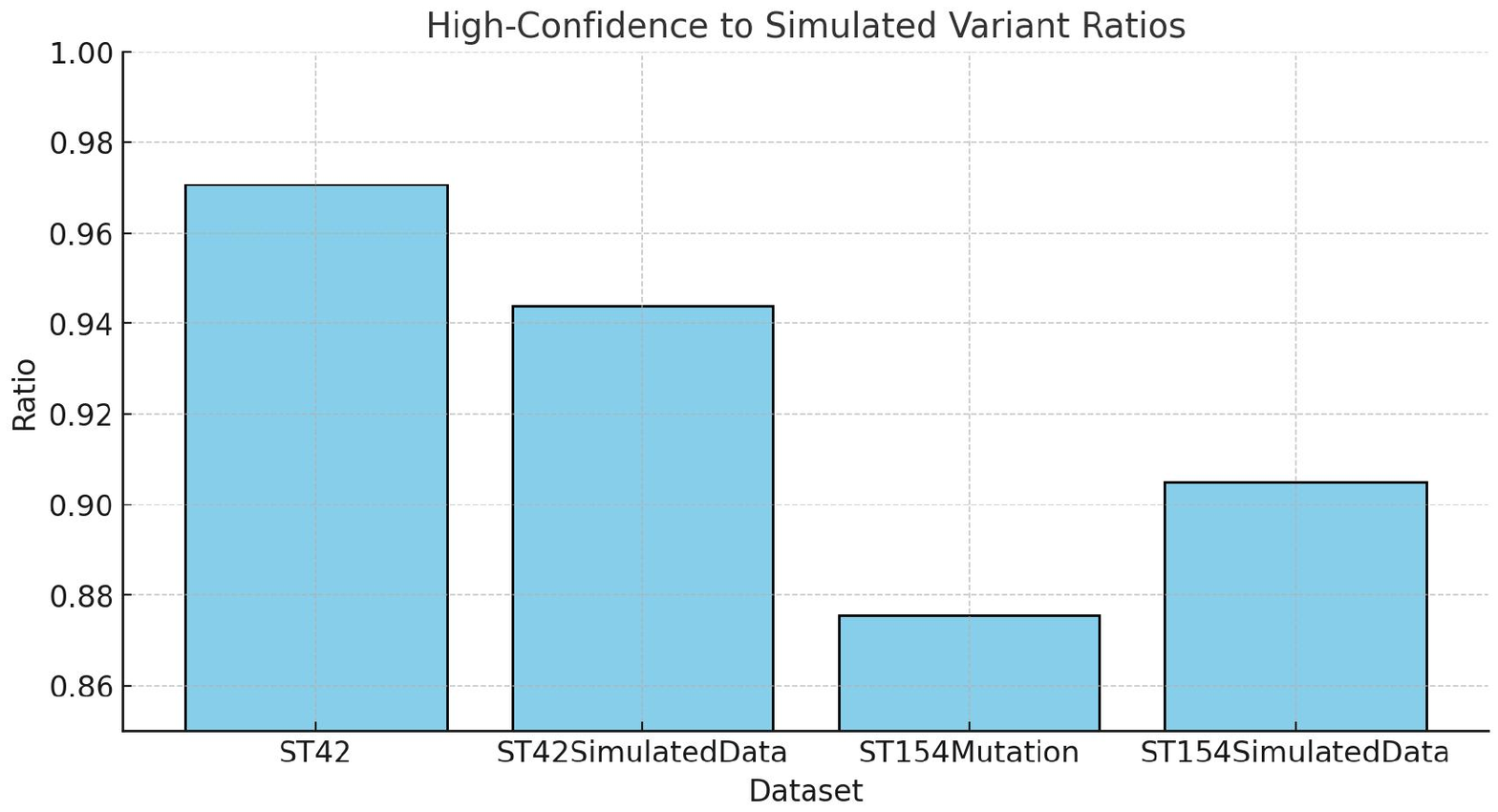}
        \caption{This bar chart illustrates the ratio of high-confidence variants to simulated variants for different datasets.}
        \label{fig:bar-chart}
    \end{subfigure}
    \begin{subfigure}[b]{0.49\linewidth}
        \centering
        \includegraphics[width=\linewidth]{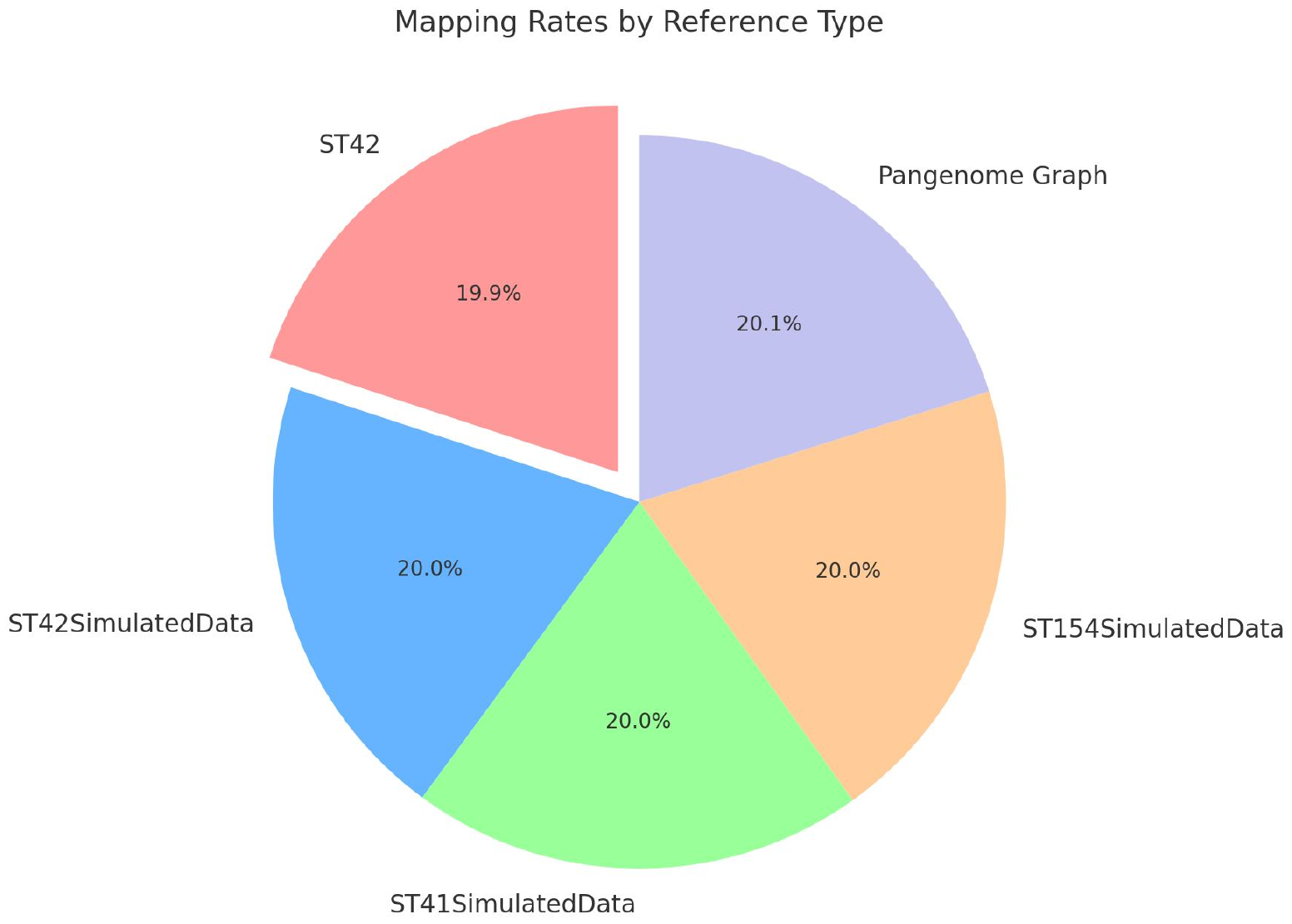}
        \caption{This is a pie chart visualising the mapping rates for different reference types, emphasising the 100\% mapping rate achieved by the pangenome graph.}
        \label{fig:pie-chart}
    \end{subfigure}
    \caption{Comparison of datasets with high-confidence variants (left) and mapping rates for reference types (right).}
    \label{fig:side-by-side}
\end{figure}










\begin{figure*}[!htbp]
\includegraphics[width=1\textwidth]{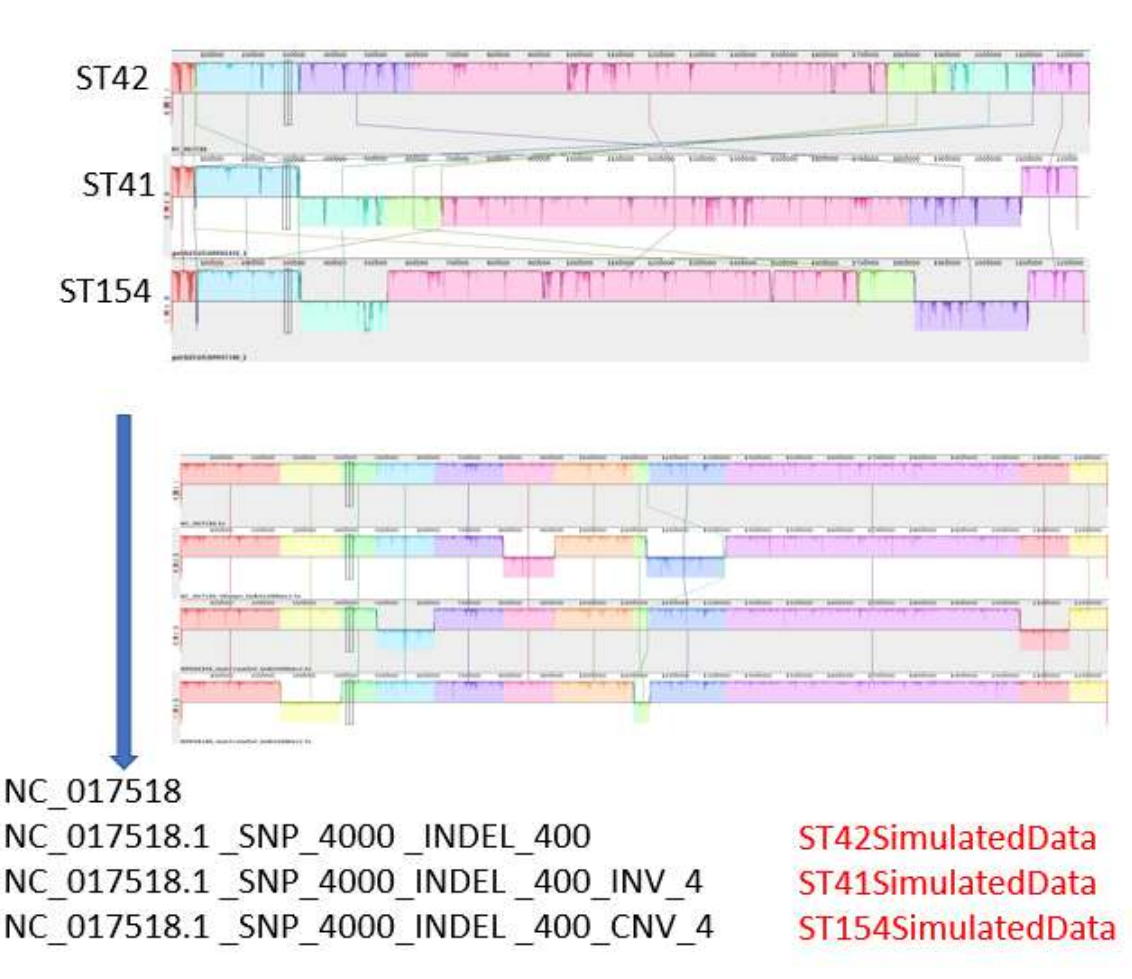}
    \caption{Simulate three genomes from ST42 reference, NC\_017518}
    \label{fig:fig6}
\end{figure*}

\section{Discussion}

Whole-genome sequencing (WGS) has revolutionised infectious disease research by enabling precise tracking of pathogen evolution, the detection of drug resistance, and facilitating vaccine development. Numerous studies \cite{Didelot:2012, Quick:2016, Gardy:2018, Geoghegan:2020, Geoghegan:2021, Yang:2021, Koser:2014, Holt:2015, Walker:2015, Chen:2021} highlight the substantial impact of WGS on public health and disease control. However, conventional linear reference-based approaches, such as bwa mem, often introduce biases, particularly when analysing highly diverse bacterial genomes \cite{Darmon:2014}. These limitations call for alternative methodologies to enhance the accuracy and inclusivity of genomic analyses.

The originality of the contribution of this study lies in its integrated and methodologically rigorous application of graph-based pangenomic tools to improve variant detection in Neisseria meningitidis, a pathogen with a highly recombinogenic genome. This study offers a thorough and repeatable bioinformatics process that uses the PanGenome Graph Builder (PGGB) and the VG toolkit (vg giraffe) to create and analyse pangenome graphs, in contrast to traditional approaches that rely on single linear reference genomes.
Pangenome graphs provide an innovative framework that encapsulates structural and minor variations—including single nucleotide polymorphisms (SNPs), insertions (INS), deletions (DEL), and inversions (INV)—within a unified graph structure. The study advances the field by (i) simulating multiple genomic datasets with diverse variant types, (ii) quantitatively comparing graph-based and linear variant calling methods, and (iii) achieving superior performance metrics for the graph-based approach in terms of sensitivity, specificity, and F1 score.

In this study, we implemented a bioinformatics pipeline leveraging the PanGenome Graph Builder (PGGB) and the Variation Graph toolbox (vg giraffe) to align WGS data, call variants against graph-based references, and construct pangenomes from assembled genomes. The pipeline consistently outperformed the linear method (bwa mem), achieving a sensitivity of 95.43\% and an F1 score of 79.98\% compared to 88.80\% and 74.46\%, respectively, for the linear approach (Section~\ref{Results}, Table~\ref{table:vg}). The paper also demonstrates the adaptability of PGGB across various genomic complexities by introducing improved parameter-tuning methodologies for building informative graphs using actual and simulated datasets. These methodological developments offer a scalable framework for comparative genomics and high-resolution pathogen surveillance while greatly reducing reference bias. Therefore, this work contributes original insights into the utility of graph-based genomics and sets a precedent for future studies in microbial variant discovery and public health surveillance.

This approach overcomes the constraints of linear references by representing the full genetic diversity of a species \cite{Paten:2017, Eizenga:2020}. Our results demonstrated that pangenome graphs improved mapping rates and enhanced variant detection accuracy for both simulated and real datasets of bacterial pathogens, supporting their utility in diverse genomic contexts.

The PGGB pipeline stands out for its ability to construct unbiased and comprehensive pangenome graphs, representing genomic diversity evenly across input genomes. This facilitated detailed analyses of Neisseria meningitidis, a pathogen with highly recombinant genomes, revealing structural variations and intricate genomic interactions. Parameter flexibility in PGGB ($-n$, $-s$, $-p$) enabled optimisation for dataset-specific complexity, with the graph exemplifying efficient construction through the $-x$ auto heuristic. These efficiencies not only reduce computational burdens but also preserve analytical precision, reassuring the practicality of adopting pangenome graphs.

Using pangenome graphs as references offers significant advantages over single linear references, reducing bias and increasing the inclusivity of genomic analyses. Graph-based workflows enable robust variant calling, including structural and copy number variations, which are critical for studying genes associated with virulence and antibiotic resistance \cite{Ekim:2021}. In our analysis, simulated datasets aligned to pangenome graph references achieved 100\% mapping rates, significantly outperforming linear references that showed mapping biases. This improved mapping and variant detection performance underscores the utility of graph-based methods for studying infectious diseases, evolutionary trends, and antibiotic resistance.

The benefits of pangenome graphs extend beyond microorganisms, with applications in human, viral, and agricultural genomics. For instance, the human draft pangenome integrates previously undetected genomic regions and variations, providing insights into complex loci and chromosomal evolution. Similarly, viral research benefits from these methods by enabling objective genomic analyses to study adaptability and guide vaccine development. The versatility of pangenome graphs opens up a world of possibilities, sparking curiosity and interest in their potential applications.

By enabling comprehensive and unbiased genomic analyses, pangenome graphs provide a robust, adaptable, and efficient tool for comparative genomics and public health research. The integration of graph-based references in genomic pipelines addresses limitations associated with linear references, improving mapping accuracy, variant detection, and the study of genetic diversity. These findings not only underscore the utility of pangenome graphs but also suggest their transformative potential, offering a new and promising approach to genomics and precision medicine.

An advantage of pangenome graphs over traditional methods is their ability to resolve complex structural rearrangements, accommodate supplementary genome content, and enable unbiased variation detection.
The framework is scalable and works with various genomic datasets, such as those from viruses, pathogens, and eukaryotic organisms. Public health implications include improved vaccine development, antibiotic resistance tracking, and genetic surveillance.


\section{Data Availability}
Reproducibility: GitHub makes the datasets and scripts used for analysis and simulation openly available: \url{https://github.com/abira-sengupta/IEEE-Women-Bioinformatics/}






\bibliographystyle{splncs04} 
\bibliography{main} 

\end{document}